\documentclass[twocolumn,english,notitlepage,superscriptaddress,nofootinbib]{revtex4-2}
\usepackage[T1]{fontenc}
\usepackage[latin9]{inputenc}
\usepackage{babel}
\usepackage{mathtools}
\usepackage{bm}
\usepackage{amsmath}
\usepackage{amssymb}
\usepackage{graphicx}

\makeatletter
\usepackage{comment}
\usepackage[T1]{fontenc}
\usepackage[latin9]{inputenc}
\usepackage{bm}
\usepackage{amsmath,amssymb}
\usepackage{graphicx}
\usepackage{xcolor}
\usepackage{mathtools,braket}
\usepackage{bbold}

\usepackage{hyperref}
\hypersetup{
    colorlinks=true,
    linkcolor=blue,
    citecolor=blue,
    filecolor=blue,      
    urlcolor=blue
     }

\makeatother

\begin{document}
\title{
Generating quantum non-local entanglement with mechanical rotations}
\author{Marko Toro\v{s}}
\affiliation{Faculty of Mathematics and Physics, University of Ljubljana, Jadranska
19, SI-1000 Ljubljana, Slovenia}
\author{Maria Chiara Braidotti }
\affiliation{School of Physics and Astronomy, University of Glasgow, Glasgow, G12
8QQ, United Kingdom}
\author{Swain Ashutosh}
\affiliation{School of Physics and Astronomy, University of Glasgow, Glasgow, G12
8QQ, United Kingdom}
\author{Mauro Paternostro}
\affiliation{Universit\`{a} degli Studi di Palermo, Dipartimento di Fisica e Chimica
- Emilio Segr\`{e}, via Archirafi 36, I-90123 Palermo, Italy}
\affiliation{Centre for Theoretical Atomic, Molecular, and Optical Physics, School
of Mathematics and Physics, Queen's University, Belfast BT7 1NN, United
Kingdom}
\author{Miles Padgett}
\affiliation{School of Physics and Astronomy, University of Glasgow, Glasgow, G12
8QQ, United Kingdom}
\author{Daniele Faccio}
\affiliation{School of Physics and Astronomy, University of Glasgow, Glasgow, G12
8QQ, United Kingdom}
\begin{abstract}
{
Recent experiments have searched for evidence of the impact of non-inertial
motion on the entanglement of particles. The success of these endeavours
has been hindered by the fact that such tests were performed within
spatial scales that were only "local" when compared to the spatial
scales over which the non-inertial motion was taking place. 
We propose a Sagnac-like interferometer that, by challenging such
bottlenecks, is able to achieve 
entangled states through a mechanism induced by the mechanical rotation
of a photonic interferometer. The resulting states violate the Bell-Clauser-Horne-Shimony-Holt
(Bell-CHSH) inequality all the way up to the Tsirelson bound, thus signaling
strong quantum nonlocality.} {Furthermore, we show that \emph{the Bell-CHSH
inequality remains violated even without using any form of post-selection} up to the value $1+\sqrt{2}$.}
Our results demonstrate that mechanical rotation can be thought of
as resource for controlling quantum non-locality {with implications
also for recent proposals for experiments that can probe the quantum
nature of curved spacetimes and non-inertial motion.} 
\end{abstract}
\maketitle

\section{Introduction }

The seminal work of J. S. Bell allowed to infer the inherent incompatibility
of quantum mechanics with the (classically acceptable) assumption
of local realism posed by Einstein, Podolsky, and Rosen~\citep{bell1964einstein,einstein1935can}.
The falsification of a Bell inequality, which would be fully satisfied
by any local realistic theory, has been reported in countless experiments~\citep{freedman1972experimental,aspect1981experimental,aspect1982experimental,weihs1998violation,pan2000experimental,rowe2001experimental,pan2012multiphoton,genovese2005research,pan2012multiphoton,brunner2014bell},
and recognised with the 2022 Nobel prize in Physics. \\
 Independently, questions about relativity led Sagnac to establish
a now widespread method for measuring rotational motion using optical
interferometry~\citep{sagnac1913preuve,sagnacEther1913}. Two counter-propagating
signals acquire a phase difference proportional to the angular frequency
of rotation~\citep{postSagnac1967,anderson1994sagnac,malykin2000sagnac,barrett2014sagnac}.
This insight led to the development of the ring laser~\citep{Macek1963}
and fiber-optical gyroscopes~~\citep{vali1976fiber,lefevre2014fiber},
with the current state-of-the-art achieving sub-shot-noise sensitivities~\citep{PhysRevLett.133.013601}.
The Sagnac effect has been shown to induce interference at the level
of quantum systems, with experimental implementations in matter-wave
interferometry~\citep{barrett2014sagnac} and single-photon platforms~\citep{bertocchi2006single}.\\
 More recently, photonic technologies have enabled the exploration
of rotation-induced quantum phenomena with two-photon experiments.
Polarization-entangled photon pairs were shown to be robust against
a 30g acceleration achieved on a rotating centrifuge~\citep{finkExperimental2017}.
Using a Hong-Ou-Mandel interferometer on a rotating platform it was
found that low frequency mechanical rotations affect bunching statistics~\citep{restuccia2019photon}.
Super-resolution and Sagnac phase sensitivity beyond the shot-noise
limit was achieved in~\citep{Fink_2019} using path-entangled NOON
states, and milli-radian phase resolution was achieved in~\citep{silvestri2024experimental},
allowing the measurement of the Earth's angular frequency of $\sim10\mu\text{Hz}$.
Furthermore, it was suggested that photonic entanglement can be revealed
or concealed using non-inertial motion accessible to current experiments~\citep{torosRevealing2020}.
Using a Hong-Ou-Mandel interferometer with nested arms it was demonstrated
that photonic behavior can change from bunching to antibunching (i.e.,
from bosonic to fermionic) solely due to mechanical rotations~\citep{cromb2023mechanical}.
Moreover, it was shown that rotational motion can change the phase
of polarization entangled states enabling transitioning between pairs
of Bell states~\citep{bittermann2024non}. {Crucially, these experiments involve quantum states that are entangled prior to rotation and rotation is only used to modify, probe, or enhance existing quantum features. \\
 A significant conceptual step} further is to demonstrate the actual generation of entanglement
using non-inertial motion. The approach proposed in Ref.~\citep{torosGeneration2022}
made use of a multi-path Sagnac interferometer to achieve a maximally
entangled path-polarization state of a single-photon. Such state would
be suitable for quantum non-contextuality tests aimed at ascertaining
whether observables can be assigned preexisting values prior to measurements
~\citep{Kochen1967}. In principle, the generated entanglement could
be also transferred to two spatially separated physical systems~\citep{van2005single},
but the experimental demonstration of such procedures remains experimentally
challenging~\citep{adhikari2010swapping,kumari2019swapping}. {It
is thus not immediately obvious whether the single-photon scheme proposed
in ~\citep{torosGeneration2022} would allow to unambiguously demonstrate
the generation of genuine quantum (nonlocal) entanglement as opposed
to local entanglement.} 
\\
 {In a different context, recent theoretical studies in quantum gravity
\citep{bose2017spin,marletto2017gravitationally} have proposed schemes
where the generation of two-particle nonlocal entanglement can be
used to witness the quantumness of the gravitational interaction mediator.
These proposals fit within a more general framework of studies where
researchers pursue methodologies to test the quantumness or nonclassicality
of the involved parties~\citep{polino2024photonic}, regardless of
the type of interaction (see also results in optomechanics and biophotonics
\citep{krisnandaRevealing2017,krisnandaPhotosyntesis2018}).} 
\\
 {Inspired by these two-party schemes,} {in this work, we propose an experimentally viable scheme - fully accessible to current photonic
technology - in which rotation itself is the mechanism that generates nonlocal entanglement between two initially unentangled photons. This marks a shift, from manipulating the entanglement to demonstrating its rotationally induced creation.}
Our scheme consists of a single Sagnac
fibre loop, linear optics elements, photon-pair sources, and a Bell-test
detection setup, all placed on a rotating platform. We find that an
initially separable two-photon state can be transformed into a maximally
entangled state of polarization that violates significantly the Bell-Clauser-Horne-Shimony-Holt
(Bell-CHSH) inequality~\citep{clauser1969proposed} as a function of the
angular frequency of rotation, obtaining a simple formula for the
frequency required to saturate Tsirelson's bound~\citep{cirel1980quantum}.
{Furthermore, we show that even \emph{without any form of post-selection},
the Bell-CHSH inequality remains violated for the same angular frequencies
of rotation, achieving the maximum value $1+\sqrt{2}$~\citep{Popescu1997}.
We conclude by briefly discussing the interpretation of the uncovered
link between mechanical motion and quantum nonlocality. } \\
%
\begin{figure*}[ht!]
\includegraphics[width=1\textwidth]{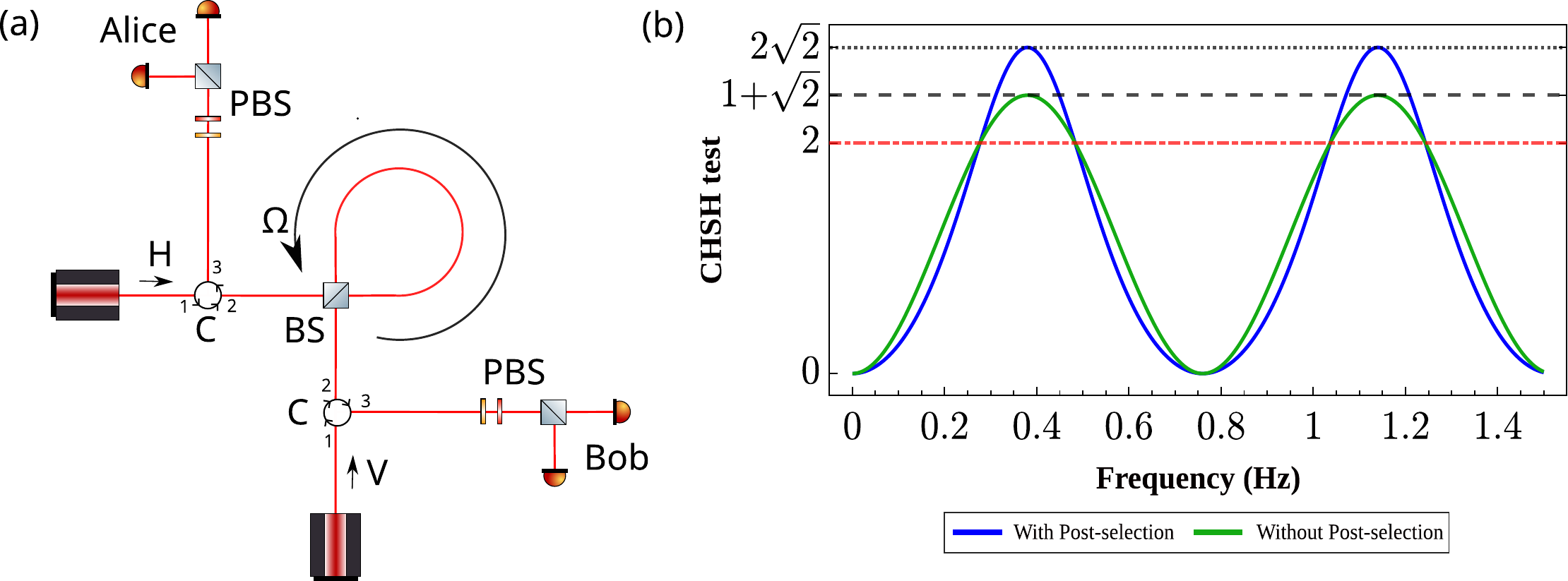} \caption{(a) Photonic setting for the 
rotation-controlled generation of quantum non-local states of polarization.
Two photons are initially prepared in the separable state $\vert HV\rangle$
and injected into the setup. {The circulators, denoted by C, first
send the photons into a Sagnac loop, and then redirect them towards
two individual detection stages (only the directional paths, $1$ to
$2$ and $2$ to $3$, are allowed by the circulators).} As the photons
entering the Sagnac loop {through Beam-splitter BS} are initially
prepared in orthogonal polarization states, they do not interact at
any point via electromagnetic couplings. We measure the polarization
of the photons with a standard Bell detection scheme. (b) Theoretical
prediction of the violation of the Bell-CHSH inequality. \emph{With
Post-selection} refers to the case where one photon is measured using
the setup at the top (managed by Alice) and one photon is measured
with the setup on the right (managed by Bob). {\emph{Without Post-selection}
refers to the case where all the photons are considered in the analysis,
which includes the detection of one photon by Alice and one of photon by Bob as well as of both photons by either Alice or Bob.}
For concreteness, we have set the photon wavelength to $\lambda=1\,\text{\ensuremath{\mu}m}$
($\omega=2\pi c/\lambda$) and the interferometric area to $\sim7.8\,\text{m}^{2}$
(e.g., $10$ loops of fiber with radius $r=0.5\,\text{m}$ with a
total length of $\sim31.5\,\text{m}$). We predict that a violation
($\vert{S}\vert>2$) occurs periodically with the rotation frequency
as stated by Eq.~(\ref{eq:omegabell}). The maximum violation is
first achieved for $\Omega_{\text{Bell}}\sim0.4\,\text{Hz}$, {and the condition $\vert S\vert>2$ is achieved in the interval $\Omega \in 0.4 \pm 0.1$ Hz (see Appendix~\ref{fluctuations} for more details).}}
\label{fig:1} 
\end{figure*}

\section{Proposed scheme.}

We consider the configuration shown in Fig.~\ref{fig:1} (a){, where two single photons are propagating through the system. The full setup involves a pulsed laser source (not shown), pumping two spatially separated type-I non-linear crystals, each producing photon pairs via spontaneous parametric down-conversion (SPDC).  One photon from each pair is detected to herald the presence of its twin, resulting in two heralded single photons that are mutually uncorrelated, as they originate from independent SPDC processes.} These signals are first directed into a Sagnac loop through {circulators and} a beam-splitter
(BS), and then towards two spatially separated
Bell-detection apparatuses. 
We assume that two independent photons are initially prepared in the
separable state: 
\begin{equation}
\vert\psi_{\text{i}}\rangle=\hat{a}_{H}^{\dagger}\hat{b}_{V}^{\dagger}\vert0\rangle\equiv\vert H\,V\rangle,\label{eq:psii}
\end{equation}
where $\hat{a}_{H}$ ($\hat{b}_{V}$) denotes horizontal H (vertical
V) polarization mode. 

As the two photons enter the Sagnac loop {through BS}, the state
changes according to a beam-splitter transformation into 
\begin{equation}
\vert\psi_{\text{i}}\rangle\rightarrow\ket{\psi_{1}}=\frac{1}{2}(\hat{a}_{H}^{\dagger}+i\hat{b}_{H}^{\dagger})(i\hat{a}_{V}^{\dagger}+\hat{b}_{V}^{\dagger})\vert0\rangle,\label{eq:2}
\end{equation}
where $\hat{a}$ ($\hat{b}$) denote the co-rotating (counter-rotating)
mode. The effect of mechanical rotation is to introduce Sagnac phases
with a sign depending on the sense of motion of the particular mode~\citep{torosRevealing2020}.
From Eq.~(\ref{eq:2}) we thus find 
\begin{equation}
\ket{\psi_{1}}{\to}\ket{\psi_{2}}{=}\frac{1}{2}\left[e^{i\frac{\phi}{2}}\hat{a}_{H}^{\dagger}{+}ie^{-i\frac{\phi}{2}}\hat{b}_{H}^{\dagger}\right]\left[ie^{i\frac{\phi}{2}}\hat{a}_{V}^{\dagger}{+}e^{-i\frac{\phi}{2}}\hat{b}_{V}^{\dagger}\right]\!\vert0\rangle,
\label{eq:3}
\end{equation}
where the phase factors have been introduced to account for the relative
phase acquired by the counter-propagating modes. Irrespective of the
medium, shape of the interferometer or the location of the center
of rotation, the Sagnac phase is given by~\citep{post1967sagnac}
\begin{equation}
\phi=\frac{4A\omega\Omega}{c^{2}},\label{eq:sagnac}
\end{equation}
where $A$ is the interferometer area, $c$ is the speed of light
in vacuum, $\omega=2\pi c/\lambda$ is the angular frequency of the
photons ($\lambda$ is the photon wavelength), $\Omega=2\pi f$, and
$f$ is the mechanical frequency of the rotating platform. Moreover,
any phase affecting differently the two polarizations would simply
factor out of Eq.~(\ref{eq:3}). Similarly, if there is some random
noise affecting the co-rotating path, then the same noise will also
affect the counter-rotating path, again factoring out of Eq.~(\ref{eq:3}).
Hence, classical phase delays arising through {experimental imperfections}
will not change the final result. To generate differential phases
depending on the direction of rotation the only plausible mechanism
is the Sagnac effect. \\
 As the photons exit the central loop {through the BS}, we apply the
inverse beam-splitter transformation to Eq.~(\ref{eq:3}), giving
us 
\begin{equation}
\begin{aligned}\ket{\psi_{2}} & {\to}\ket{\psi_{\text{tot}}}{=}\\
 & =\frac{1}{4}\left[\left(e^{i\phi/2}(-i\hat{a}_{H}^{\dagger}+\hat{b}_{H}^{\dagger})+ie^{-i\phi/2}(\hat{a}_{H}^{\dagger}-i\hat{b}_{H}^{\dagger})\right)\right.\\
 & \times\left.\left(ie^{i\phi/2}(-i\hat{a}_{V}^{\dagger}+\hat{b}_{V}^{\dagger})+e^{-i\phi/2}(\hat{a}_{V}^{\dagger}-i\hat{b}_{V}^{\dagger}\right)\right]\vert0\rangle.
\end{aligned}
\label{eq:5}
\end{equation}
{After rearranging the terms in Eq.~\eqref{eq:5} we find
\begin{alignat}{1}
\vert\psi_{\text{tot}}\rangle= & \frac{1}{2}\sqrt{3+\text{cos}(2\phi)}\vert\psi_{\text{\text{\text{fav}}}}\rangle+\frac{\text{sin}\phi}{\sqrt{2}}\vert\psi_{\text{\text{\text{unfav}}}}\rangle,\label{eq:tot}
\end{alignat}
where we have introduced the normalized states

\begin{alignat}{1}
&\vert\psi_{\text{\text{\text{fav}}}}\rangle  =\left[\frac{(\text{cos}\phi-1)\hat{a}_{H}^{\dagger}\hat{b}_{V}^{\dagger}}{\sqrt{3+\text{cos}(2\phi)}}+\frac{(\text{cos}\phi+1)\hat{b}_{H}^{\dagger}\hat{a}_{V}^{\dagger}}{\sqrt{3+\text{cos}(2\phi)}}\right]\vert0\rangle,\label{eq:fav}\\
&\vert\psi_{\text{\text{unfav}}}\rangle  =\frac{1}{\sqrt{2}}\left[\hat{a}_{H}^{\dagger}\hat{a}_{V}^{\dagger}-\hat{b}_{H}^{\dagger}\hat{b}_{V}^{\dagger}\right]\vert0\rangle.\label{eq:unfav}
\end{alignat}
$\vert\psi_{\text{\text{\text{fav}}}}\rangle$ represents the case when
each pair of detectors detects one photon, i.e., Alice detects the
mode $a$ and Bob detects mode $b$ or vice versa, while $\vert\psi_{\text{\text{\text{unfav}}}}\rangle$
represents the case when both photons arrive at the same pair of detectors,
i.e., either both to Alice or both to Bob. As we want to compute quantum
correlations between Alice and Bob we will refer to $\vert\psi_{\text{\text{\text{fav}}}}\rangle$
and $\vert\psi_{\text{\text{\text{unfav}}}}\rangle$ as the \emph{favorable}
and \emph{unfavorable} state, respectively. }

\section{Violation of the Bell-CHSH inequality}

To quantify the degree of generated non-locality we can compute the
Bell-CHSH function $S$~\citep{clauser1969proposed} 
\begin{equation}
S=E(\bm{a},\bm{b})-E(\bm{a},\bm{b}')+E(\bm{a}',\bm{b})+E(\bm{a}',\bm{b}'),\label{eq:S}
\end{equation}
where the quantum correlations are given by

\begin{equation}
E(\bm{v},\bm{u})=\langle\psi\vert\hat{A}_{\bm{v}}\hat{B}_{\bm{u}}\vert\psi\rangle,
\end{equation}
{$\vert\psi\rangle$ is the state of the system, and $\hat{A}_{\bm{v}}$
and $\hat{B}_{\bm{u}}$ correspond to the measurements performed by
Alice and Bob, respectively ($\bm{v}=\bm{a}$,$\bm{a}'$ and $\bm{u}=\bm{b}$,$\bm{b}'$
denote unit vectors parametrizing the measurement operators). The
four terms appearing in Eq.~\eqref{eq:S} are assocated to four different
joint measurements performed by Alice and Bob. 
}
Local realism enforces the Bell-CHSH inequality $\vert S\vert\leq2$.
However, as we will see, suitable choices of the angular frequency
of rotation, encoded in the phase $\phi$ in Eq.~\eqref{eq:tot},
allows to violate such constraint. {We show below that the Bell-CHSH
inequality is violated up to the value $\vert S\vert=1+\sqrt{2}$
by considering the total state in Eq.~(\ref{eq:tot}), establishing
the phenomena of rotationally induced quantum non-locality (Sec.~\ref{wops}).} Furthermore,
we show that Tsirelson's bound $\vert S\vert=2\sqrt{2}$ can be achieved
by post-selecting coincidences at the detectors (Sec.~\ref{wps}).

\subsection{Non-locality without post-selection}\label{wops}

{We define the observables $\hat{A}_{\bm{u}}$ and $\hat{B}_{\bm{v}}$ by prescribing how they act on the basis
states of our Hilbert space (we recall that observables, which are linear maps, can always be defined in this way). The total Hilbert space, and in particular its basis, can be read out 
from the total final state $\vert\psi_\text{tot}\rangle$ in Eqs.~\eqref{eq:tot}-\eqref{eq:unfav}. On the Hilbert subspace spanned by $\hat{a}_{H}^{\dagger}\hat{b}_{V}^{\dagger}\vert0\rangle$
and $\hat{b}_{H}^{\dagger}\hat{a}_{V}^{\dagger}\vert0\rangle$ (i.e.,
the basis states of the favourable subspace) we define $\hat{A}_{\bm{a}}\equiv\bm{a}\cdot\bm{\sigma}$
and $\hat{B}_{\bm{b}}\equiv\bm{b}\cdot\bm{\sigma}$, where $\bm{\sigma}=(\sigma_{x},\sigma_{y},\sigma_{z})$
is the vector of the Pauli matrices. In order to achieve the highest
violation of the Bell test, we choose the vectors $\bm{a}=(1,0,0)$,
$\bm{a}'=(0,1,0)$, $\bm{b}=(1,1,0)/\sqrt{2}$, $\bm{b}'=(-1,1,0)/\sqrt{2}$.
On the Hilbert subspace spanned by $\hat{a}_{H}^{\dagger}\hat{a}_{V}^{\dagger}\vert0\rangle$
and $\hat{b}_{H}^{\dagger}\hat{b}_{V}^{\dagger}\vert0\rangle$ (i.e.,
the basis states of the unfavourable subspace) we define $\hat{A}_{\bm{a}}=\mathbb{I}$
and $\hat{B}_{\bm{a}}=\mathbb{I}$, where $\mathbb{I}$ denotes the
identity operator~\citep{Popescu1997}. 

To compute the Bell-CHSH value we consider the total state without any post-selection, i.e.,
we set $\vert\psi\rangle\equiv\vert\psi_{\text{tot}}\rangle$. Inserting
Eq.~(\ref{eq:tot}) into Eqs.~(\ref{eq:S}) we eventually find:
\begin{equation}
S=\left(1+\sqrt{2}\right)\sin^{2}(\phi).\label{eq:Svalue}
\end{equation}
In particular, by setting $\phi=\pi/2+k\pi~(k\in\mathcal{Z})$, the
state in Eq.~(\ref{eq:tot}) reduces to $\vert\psi_{\text{tot}}\rangle=\frac{1}{\sqrt{2}}(\psi_{\text{\text{\text{fav}}}}\rangle+\vert\psi_{\text{\text{\text{unfav}}}}\rangle)$,
i.e.,
\begin{equation}
\vert\psi_{\text{tot}}\rangle=\frac{1}{2}\left[-\hat{a}_{H}^{\dagger}\hat{b}_{V}^{\dagger}+\hat{b}_{H}^{\dagger}\hat{a}_{V}^{\dagger}+\hat{a}_{H}^{\dagger}\hat{a}_{V}^{\dagger}-\hat{b}_{H}^{\dagger}\hat{b}_{V}^{\dagger}\right]\vert0\rangle,\label{eq:totalCHSH}
\end{equation}
 and we achieve the maximum violation $\vert S\vert=1+\sqrt{2}$.
This occurs when the mechanical frequency $\Omega$ takes the values
\begin{equation}
\Omega_{\text{Bell}}\equiv\frac{\pi c^{2}}{8A\omega}(2k+1),\qquad(k\in\mathcal{Z}).\label{eq:omegabell}
\end{equation}
Here, $k<0$ ($k>0$) would correspond to an (anti-) clockwise sense
of rotation.

The experimentally measurable Bell-CHSH correlation will be obtained by
repeating the experiment with a large number of initial photon pairs
in order to gather enough statistics. In Fig.~\ref{fig:1}(b), we
plot the Bell-CHSH function for a set of values of the relevant physical
parameters that are well within reach of existing photonic technology with previous experiments already achieving the required sensitivities~\citep{finkExperimental2017,torosRevealing2020,cromb2023mechanical,Fink_2019,silvestri2024experimental,bittermann2024non,restuccia2019photon}.

Importantly, the Bell-CHSH violation depends critically only on the mechanical rotation.
We recall that two initial state in Eq.~\eqref{eq:psii} consists of photons that are in orthogonal polarization modes,
and hence do not interact at the beam-splitter or via any electromagnetic interaction, 
and that no post-selection has been performed on the total final state in Eq.~\eqref{eq:tot}. 
The results captured by Eq.~(\ref{eq:S}) thus depend critically only
on the angular frequency of rotation $\Omega$: by tuning its value we can generate
quantum non-locality, while without mechanical rotation no quantum
non-locality can be established. }

\subsection{Tsirelson's bound with postselection}\label{wps}

{We now consider the state conditional on the postselection of the
events that provide coincidences at the detectors, i.e., quantum correlations
between Alice and Bob, where we disregard the events arising from
the unfavorable state $\vert\psi_{\text{unfav}}\rangle$ in Eq.~\eqref{eq:unfav}. 
We discuss the probability of coincidence detection in Appendix~\ref{probability}, which remains greater than $50\%$, and hence does not pose a severe limitation.}
We can rewrite the remaining favorable state $\vert\psi_{\text{fav}}\rangle$
defined in Eq.~\eqref{eq:fav} as: 
\begin{alignat}{1}
\vert\psi_{\text{fav}}\rangle & =\frac{\text{cos}(\phi)+1}{\sqrt{3+\text{cos}(2\phi)}}\vert H\,V\rangle+\frac{\text{cos}(\phi)-1}{\sqrt{3+\text{cos}(2\phi)}}\vert V\,H\rangle,\label{eq:psif}
\end{alignat}
where we have introduced the commonly used notation $\hat{a}_{H}^{\dagger}\hat{b}_{V}^{\dagger}\vert0\rangle\equiv\vert H\,V\rangle$
and $\hat{a}_{V}^{\dagger}\hat{b}_{H}^{\dagger}\vert0\rangle\equiv\vert V\,H\rangle$. 

We first note that for $\phi=0$ (corresponding to the case without
mechanical rotation) we always remain in the initial state $\vert H\,V\rangle$,
which is separable. More generally, we note that for $\phi=\pi\,k$
($k\in\mathcal{Z})$ we remain either in the initial state $\vert H\,V\rangle$
( {$k$} even) or transform into the flipped polarization state
{$\vert V\,H\rangle$} 
( {$k$} odd). However, for any other value of $\phi$ we find that
Eq.~(\ref{eq:psif}) will be in an entangled state. In particular,
for $\phi=\pi/2+\pi\,k$ ($k\in\mathcal{Z})$ Eq.~(\ref{eq:psif})
transforms into the maximally entangled Bell state 
\begin{equation}
\vert\psi_{\text{fav}}\rangle=\frac{1}{\sqrt{2}}\left(\vert H\,V\rangle-\vert V\,H\rangle\right),\label{eq:psifBell}
\end{equation}
which is usually denoted as the $\vert\Psi^{-}\rangle$ state.

To compute the Bell-CHSH value we set $\vert\psi\rangle\equiv\vert\psi_{\text{f}}\rangle$
in Eq.~(\ref{eq:S}) to obtain the value

\begin{equation}
S=4\sqrt{2}\frac{\sin^{2}(\phi)}{3+\cos(2\phi)}.\label{eq:SvaluePS}
\end{equation}
In particular, as stated in Eq.~(\ref{eq:psifBell}), by setting
$\phi=\pi/2+k\pi~(k\in\mathcal{Z})$, the state in Eq.~(\ref{eq:psif})
reduces to the Bell state $\vert\Psi^{-}\rangle$ and we achieve the
notorious Tsirelson's bound $\vert S\vert=2\sqrt{2}$~\citep{cirel1980quantum}.
In Fig.~\ref{fig:1}(b), we plot the Bell-CHSH function defined in Eq.~\eqref{eq:SvaluePS} for the same set of values as for the case without postselection to ease the comparison. 

Saturating Tsirelson's bound depends critically on two steps: (i) on the post-selection
step from Eq.~(\ref{eq:tot}) to (\ref{eq:psif}), where we have
discarded the  unfavorable state, and (ii)
on a non-zero rotationally induced phase $\phi\neq0$, which arises
only for non-zero frequencies of rotation $\Omega\neq0$. The post-selection step is however not enough to induce entanglement in the
absence of mechanical rotation as discussed above for the case $\phi=0$.
{Furthermore, even without any form post-selection we have shown that the Bell-CHSH inequality remains violated (see Sec.~\ref{wops}).}

\section{Discussion}

We have proposed a method for the controlled generation of quantum
non-locality using mechanical rotation that achieves the map
\begin{alignat}{1}
\hat{a}_{H}^{\dagger}\hat{b}_{V}^{\dagger}\vert0\rangle\rightarrow\frac{1}{2}\bigg[ & (\text{cos}\phi-1)\hat{a}_{H}^{\dagger}\hat{b}_{V}^{\dagger}+(\text{cos}\phi+1)\hat{b}_{H}^{\dagger}\hat{a}_{V}^{\dagger}\nonumber \\
 & +\text{sin\ensuremath{\phi}}\,(\hat{a}_{H}^{\dagger}\hat{a}_{V}^{\dagger}-\hat{b}_{H}^{\dagger}\hat{b}_{V}^{\dagger})\bigg]\vert0\rangle.\label{eq:finalEq}
\end{alignat}
The minimalist derivation is fully contained in Eqs.~\eqref{eq:psii}-\eqref{eq:unfav}, assuming only the Sagnac phase and the form of the beam-splitter transformation.
By controlling
the angular frequency of the mechanical rotation, $\Omega$, and hence
the corresponding Sagnac phase, $\phi\equiv\phi(\Omega)$, we have
shown that we can either prepare separable or nonlocally entangled
final states. 

The map in Eq.~\eqref{eq:finalEq} satisfies a number of desiderata: (i) For $\phi=0$
the transformation reduces to the identity map, i.e., $\mathcal{M}=\mathbb{I}$. In other words, without
mechanical rotation, the state remains invariant (and classical).
(ii) For $\phi=\pi\,k$ ($k\in{\mathbb{Z}})$ the transformation in
Eq.~\eqref{eq:finalEq} is either the identity operation (even $k$)
or induces a polarization flip (odd $k$). The latter case shows that
mechanical rotation can be used to swap the polarization state of
photon pairs. (iii) For $\phi=(2k+1)\pi/2$ ($k\in{\mathbb{Z}})$
we obtain the state in Eq.~\eqref{eq:totalCHSH}, which is predicted
to violate the Bell-CHSH inequality. The first line of Eq.~\eqref{eq:finalEq}
(corresponding to Alice and Bob detecting each one photon) reduces to the Bell state
$\vert\Psi^{-}\rangle$, which is expected to induce a maximal violation
of the Bell-CHSH inequality given by the Tsirelson's bound~\citep{cirel1980quantum}.
{However, even in absence of any form of post-selection, the total final state
in Eq.~\eqref{eq:finalEq} produces a violatation of the Bell-CHSH inequality up to the value $S=1+\sqrt{2}$~\cite{Popescu1997}.} 

The scheme is also robust against imperfections and noise due to the
inherent protection characteristic of the Sagnac loop. Suppose some
unwanted phases would be accumulating depending on the polarization
$H$, $V$; this would contribute only to a global phase in Eq.~(\ref{eq:3}),
but no measurable differential phase would be generated. Similarly,
any other random phase affecting the co-rotating path will automatically
affect also the counter-rotating path, thus factoring out without
affecting the final state. As shown in Fig.~\ref{fig:1}(b) the experimental
parameters required to test the maximum violation of the Bell-CHSH inequality
can be achieved with current photonic technologies by adaptation of
previous experimental schemes~\citep{finkExperimental2017,torosRevealing2020,cromb2023mechanical,Fink_2019,silvestri2024experimental,bittermann2024non,restuccia2019photon}.
As such we do not expect any new fundamental or technical issue in
the implementation of this proposal (see Appendix~\ref{noise}
for the analysis of background noise, dark counts, and detector
inefficiencies). 


A further benefit of the proposed scheme is also that it does not
rely on specific models, but rather on the well established Bell-CHSH test.
The violation arises only when we tune the mechanical frequency of
rotation to the interval centered on the value $\Omega_{\text{Bell}}$. Hence
non-zero mechanical rotation is a critical factor for generating non-locality
in this setup, i.e., we can {legitimately} speak of \emph{rotationally
induced nonlocality}.

The question of how to interpret the experiment is of course nonetheless
interesting. In this work we have provided a simple yet very effective
and powerful theoretical interpretation only relying on the Sagnac
phase. While here we have not shown this, the Sagnac phase is of intrinsic
relativistic origin. Evidence of such nature stems from Eq.~(\ref{eq:sagnac}),
which depends on the speed of light in vacuum and not on that of photons
in a medium, suggesting that its origin is related to the spacetime
metric (we refer the interested reader to the reviews~\citep{postSagnac1967,anderson1994sagnac,malykin2000sagnac,barrett2014sagnac}).
However, more formal interpretations within quantum theory in curved
space~\citep{torosRevealing2020}, broader quantum field theoretic
framework~\citep{korsbakken2004fulling}, or a general relativistic
context~\citep{zych2012general,brady2021frame,HOM_Kerr_2022,barzelObserver2022}
are also possible. The possibility to further such thoughts and 
interpret the spacetime metric as in a superposition, along the lines
of Ref.~\citep{torosGeneration2022}, thus reaching out to the domain
of quantum reference~\citep{aharonov1984quantum,giacomini2019relativistic}
frames, will be the topic of further investigations.
%

\subsection*{Acknowledgements}

MT acknowledges funding from the Slovenian Research and Innovation
Agency (ARIS) under contracts N1-0392, P1-0416, SN-ZRD/22-27/0510
(RSUL Toro\v{s}) and from the Leverhulme Trust (RPG-2020-197). MP
acknowledges the support by the Horizon Europe EIC Pathfinder project
QuCoM (Grant Agreement No.\,101046973), the Leverhulme Trust Research
Project Grant UltraQuTe (grant RGP-2018-266), the Royal Society Wolfson
Fellowship (RSWF/R3/183013), the UK EPSRC (EP/T028424/1), and the
Department for the Economy Northern Ireland under the US-Ireland R\&D
Partnership Programme. DF acknowledges support from the Royal Academy
of Engineering and the UK EPSRC (EP/W007444/1).

\appendix

\section{Angular velocity fluctuations}\label{fluctuations}

{The presented scheme is robust with respect to angular velocity
fluctuations, $\delta\Omega$, of the rotating platform. For example,
let us focus on the region of the first peak shown in Fig.~\ref{fig:1},
where $\phi\in[0,\pi]$. We can then readily estimate the minimum
(maximum) angular frequency $\Omega_{-}$ ($\Omega_{+}$) at the boundaries
of the interval, where we have a violation of the Bell inequality,
i.e., determined by the condition $\vert S\vert>2$. We define the
maximum allowed angular velocity fluctuations as $\delta\Omega\equiv\Omega_{+}-\Omega_{\text{Bell}}$
(or equivalently $\delta\Omega=\Omega_{\text{Bell}}-\Omega_{-}$ given
the the symmetric shape of the peak), such that for angular
frequencies in the interval
\begin{equation}
\Omega\in\Omega_{\text{Bell}}\pm\delta\Omega\label{eq:interval}
\end{equation}
 we have a violation of the Bell inequality. From Eqs.~\eqref{eq:sagnac}
and ~\eqref{eq:Svalue} (or from Eqs.~\eqref{eq:sagnac} and \eqref{eq:SvaluePS}), we find the that the maximum allowed angular
velocity fluctuations are given by
\begin{equation}
\delta\Omega=\frac{c^{2}}{4A\omega}\arctan\sqrt{\frac{\sqrt{2}-1}{2}}.
\end{equation}
In particular, for the parameters listed in Fig.~\ref{fig:1} we
find the value $\delta\Omega=2\pi\times0.1\text{Hz}$, which does
not pose a challenging requirement for the rotation control of low-frequency rotating platforms~\citep{restuccia2019photon,cromb2023mechanical}.
A similar analysis on the sensitivity of the Bell-CHSH violation can be
performed also for the other peaks at higher ($k>0$) or negative
($k<0$) values of $\Omega_{\text{Bell}}$.}

\section{Coincidence detection}\label{probability}
{The detection method considered here relies on four-photon coincidences: two photon pairs are generated by two different SPDC processes. One photon from each pair is heralded (not shown in Fig.~\ref{fig:1}(a)), ensuring that the other photon propagates within the system. In this frame, the loss of one photon do not affect the measurement; it will only result in longer acquisition times. In the following discussion, we do not explicitly include the heralding process; all references to two-photon detection or coincidence events are made under the implicit assumption that the corresponding heralded photons have been successfully detected.} \\ 
{A further point needs to be made: in the case with post-selection, not all two-photons detections produce an increase in the number of counts; only when the top-left pair of detectors
(Alice) detects one photon and the bottom-right pair of detectors
(Bob) detects one photon we are able to update the experimental value
of $S$ defined in Eq.~\eqref{eq:SvaluePS} (see Fig.~\ref{fig:1}(a)) - we need a coincidence measurement between Alice and Bob. We can find the resulting reduction
in the number of counts by noting that the numerical prefactors of Eq.~(\ref{eq:tot}) and Eq.~(\ref{eq:fav}) give the associated
probability amplitudes. Hence, squaring and summing the amplitudes
of the terms $\propto\hat{a}^{\dagger}\hat{b}^{\dagger}$ we find
the probability of detection $(\text{cos}^{2}(\phi)+1)/2$, which
gives a $50\%$ reduction in the statistics at $\Omega=\Omega_{\text{Bell}}.$
The case without post-selection is unaffected by this reduction in
the probability of detection as all the photons are taken into account.}

\section{Effect of Noise on Bell-CHSH Test}\label{noise}

{\begin{figure*}[t]
\includegraphics[width=1\textwidth]{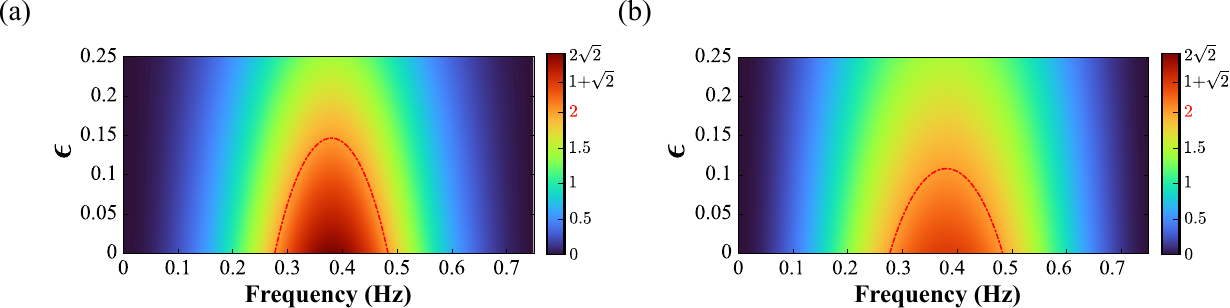} \caption{Variation of the Bell-CHSH parameter S with frequency and error probability
$\epsilon$ (a) Case with post-selection (b) Case without post-selection.
We use same values for wavelength $\lambda$ and interferometric area
as used in Fig.~\ref{fig:1}.}
\label{fig:2}
\end{figure*}

Noise in the form of background noise, dark counts, and detector inefficiencies
can significantly affect the quantum correlations relevant to the
Bell-CHSH test. In this section, we quantitatively examine the effect
of such noise on the Bell-CHSH test and consequently on the extent
of violation of local-realism. Such investigations are of paramount
importance when trying to assess the feasibility of our scheme for
experimental implementation.\\}
{One possible noise source is a passive coupling mechanism caused by back-scattering from the cavity mirrors, in particular the BS surfaces in our setup. While such coupling could, in principle, lead to spurious mode interactions or background contributions, it is not expected to play a significant role in our case. The single photons do not populate a well-defined intracavity mode, and therefore do not establish conditions for mode-locking through mirror-induced feedback. Furthermore, due to the low generation efficiencies, as well as the finite spectral bandwidth of the photons, their coherence length remains short. This prevents significant temporal or spatial overlap between co-rotating and counter-rotating photon wave packets, effectively suppressing coherent interference effects arising from back-scattering.\\
Low pair-generation efficiencies also play a crucial role in suppressing unwanted multi-pair emission events, which can introduce noise into the measurement outcomes. Similar to what occurs in static (non-rotating) measurements, such multi-pair events contribute to a small but non-negligible background. However, their impact can be minimized by operating at low pump powers, reducing the probability of higher-order emissions.\\ 
A separate treatment is necessary when considering a noise model in which the subsystems associated with Alice and Bob are subject to independent error processes with noise error probability $\epsilon_{A}$ and $\epsilon_{B}$, respectively
\citep{Brunner2007}.} { To simplify the presentation and estimate the
magnitude of the effects, we here assume that the noise error probability
of Alice's subsystem, $\epsilon_{A}$, is the same as Bob's, $\epsilon_{B}$
($\epsilon_{A}=\epsilon_{B}=\epsilon$), and that $0<\epsilon\ll1$.
Hence, we can write the total system statistical operator $\hat{\rho}$
as:

\begin{equation}
\hat{\rho}=(1-2\epsilon)\hat{\rho}_{AB}+\epsilon\left(\hat{\rho}_{A}\otimes\frac{1}{2}\right)+\epsilon\left(\frac{1}{2}\otimes\hat{\rho}_{B}\right),\label{eq:N}
\end{equation}

\noindent where $\hat{\rho}_{AB}=\ket{\psi_{3}}\bra{\psi_{3}}$, with
$\ket{\psi_{3}}$ the interferometer output state give by Eq.~(\ref{eq:tot}),
while $\hat{\rho}_{A}=\text{tr}_{B}\left(\ket{\psi_{3}}\bra{\psi_{3}}\right)$
and $\hat{\rho}_{B}=\text{tr}_{A}\left(\ket{\psi_{3}}\bra{\psi_{3}}\right)$
are the reduced states of Alice's and Bob's subsystem. We use $\hat{\rho}$
to compute the Bell-CHSH function $S$ from Eq.~(\ref{eq:S}) for
two cases: without post-selection (Sec.~\ref{wops}) and with post-selection
(Sec.~\ref{wps}).\\
 To evaluate E(a,b) in $S$, we use: 
\begin{equation}
E(\bm{a},\bm{b})=\text{tr}(\hat{O}\,\hat{\rho}),
\end{equation}
where $\hat{O}$ is equal to $(\bm{a}\cdot\bm{\sigma})\otimes(\bm{b}\cdot\bm{\sigma})$
for the case with post-selection (we have one photon for
each subsystem). However, for the case without post-selection, we use
the approach suggested in \citep{Popescu1997}: here $\hat{O}$ is
equal to $(\bm{a}\cdot\bm{\sigma})\otimes(\bm{b}\cdot\bm{\sigma})$
when each subsystem receives one photon, while $\hat{O}$ is equal
to identity for the other cases when Alice or Bob's subsystem receive
more than one photon each.\\
Using this formalism, for the case with post-selection, we get:
\begin{equation}
S=(1-2\epsilon)4\sqrt{2}\frac{\sin^{2}\phi}{3+\cos{2\phi}}.
\end{equation}
Similarly for the case without post-selection, we get:
\begin{equation}
  S=(1  -2\epsilon)(1+\sqrt{2)}\sin^{2}\phi+\epsilon\sin^{2}{\phi}.\label{eq25}  
\end{equation}

\noindent The variation of $S$ with frequency and the error probability $\epsilon$
is shown in Fig. \ref{fig:2}. As expected, for the case with post-selection,
we see a violation of the Bell-CHSH inequality for a higher error
probability compared to the case without post-selection. Nevertheless, it is worth
noticing that the Bell-CHSH parameter can exceed the threshold value $S=2$ in both cases. 
This is due to the fact that $S$ measures
the nonlocality of the state and hence that our final state is indeed
nonlocal, even if not maximally entangled.}


\begin{thebibliography}{0}%
\makeatletter
\providecommand \@ifxundefined [1]{%
 \@ifx{#1\undefined}
}%
\providecommand \@ifnum [1]{%
 \ifnum #1\expandafter \@firstoftwo
 \else \expandafter \@secondoftwo
 \fi
}%
\providecommand \@ifx [1]{%
 \ifx #1\expandafter \@firstoftwo
 \else \expandafter \@secondoftwo
 \fi
}%
\providecommand \natexlab [1]{#1}%
\providecommand \enquote  [1]{``#1''}%
\providecommand \bibnamefont  [1]{#1}%
\providecommand \bibfnamefont [1]{#1}%
\providecommand \citenamefont [1]{#1}%
\providecommand \href@noop [0]{\@secondoftwo}%
\providecommand \href [0]{\begingroup \@sanitize@url \@href}%
\providecommand \@href[1]{\@@startlink{#1}\@@href}%
\providecommand \@@href[1]{\endgroup#1\@@endlink}%
\providecommand \@sanitize@url [0]{\catcode `\\12\catcode `\$12\catcode `\&12\catcode `\#12\catcode `\^12\catcode `\_12\catcode `\%12\relax}%
\providecommand \@@startlink[1]{}%
\providecommand \@@endlink[0]{}%
\providecommand \url  [0]{\begingroup\@sanitize@url \@url }%
\providecommand \@url [1]{\endgroup\@href {#1}{\urlprefix }}%
\providecommand \urlprefix  [0]{URL }%
\providecommand \Eprint [0]{\href }%
\providecommand \doibase [0]{https://doi.org/}%
\providecommand \selectlanguage [0]{\@gobble}%
\providecommand \bibinfo  [0]{\@secondoftwo}%
\providecommand \bibfield  [0]{\@secondoftwo}%
\providecommand \translation [1]{[#1]}%
\providecommand \BibitemOpen [0]{}%
\providecommand \bibitemStop [0]{}%
\providecommand \bibitemNoStop [0]{.\EOS\space}%
\providecommand \EOS [0]{\spacefactor3000\relax}%
\providecommand \BibitemShut  [1]{\csname bibitem#1\endcsname}%
\let\auto@bib@innerbib\@empty
\end{thebibliography}%


\begin{thebibliography}{10}

\bibitem{bell1964einstein}
John~S Bell.
\newblock On the einstein podolsky rosen paradox.
\newblock {\em Physics Physique Fizika}, 1(3):195, 1964.

\bibitem{einstein1935can}
Albert Einstein, Boris Podolsky, and Nathan Rosen.
\newblock Can quantum-mechanical description of physical reality be considered complete?
\newblock {\em Physical review}, 47(10):777, 1935.

\bibitem{freedman1972experimental}
Stuart~J Freedman and John~F Clauser.
\newblock Experimental test of local hidden-variable theories.
\newblock {\em Physical Review Letters}, 28(14):938, 1972.

\bibitem{aspect1981experimental}
Alain Aspect, Philippe Grangier, and G{\'e}rard Roger.
\newblock Experimental tests of realistic local theories via bell's theorem.
\newblock {\em Physical review letters}, 47(7):460, 1981.

\bibitem{aspect1982experimental}
Alain Aspect, Jean Dalibard, and G{\'e}rard Roger.
\newblock Experimental test of bell's inequalities using time-varying analyzers.
\newblock {\em Physical review letters}, 49(25):1804, 1982.

\bibitem{weihs1998violation}
Gregor Weihs, Thomas Jennewein, Christoph Simon, Harald Weinfurter, and Anton Zeilinger.
\newblock Violation of bell's inequality under strict einstein locality conditions.
\newblock {\em Physical Review Letters}, 81(23):5039, 1998.

\bibitem{pan2000experimental}
Jian-Wei Pan, Dik Bouwmeester, Matthew Daniell, Harald Weinfurter, and Anton Zeilinger.
\newblock Experimental test of quantum nonlocality in three-photon greenberger--horne--zeilinger entanglement.
\newblock {\em Nature}, 403(6769):515--519, 2000.

\bibitem{rowe2001experimental}
Mary~A Rowe, David Kielpinski, Volker Meyer, Charles~A Sackett, Wayne~M Itano, Christopher Monroe, and David~J Wineland.
\newblock Experimental violation of a bell's inequality with efficient detection.
\newblock {\em Nature}, 409(6822):791--794, 2001.

\bibitem{pan2012multiphoton}
Jian-Wei Pan, Zeng-Bing Chen, Chao-Yang Lu, Harald Weinfurter, Anton Zeilinger, and Marek {\.Z}ukowski.
\newblock Multiphoton entanglement and interferometry.
\newblock {\em Reviews of Modern Physics}, 84(2):777, 2012.

\bibitem{genovese2005research}
Marco Genovese.
\newblock Research on hidden variable theories: A review of recent progresses.
\newblock {\em Physics Reports}, 413(6):319--396, 2005.

\bibitem{brunner2014bell}
Nicolas Brunner, Daniel Cavalcanti, Stefano Pironio, Valerio Scarani, and Stephanie Wehner.
\newblock Bell nonlocality.
\newblock {\em Reviews of modern physics}, 86(2):419, 2014.

\bibitem{sagnac1913preuve}
Georges Sagnac.
\newblock Sur la preuve de la r{\'e}alit{\'e} de l'{\'e}ther lumineux par l'exp{\'e}rience de l'interf{\'e}rographe tournant.
\newblock {\em CR Acad. Sci.}, 157:1410--1413, 1913.

\bibitem{sagnacEther1913}
Georges Sagnac.
\newblock L'\'ether lumineux d\'emontr\'e.
\newblock {\em Comptes rendus hebdomadaires des s\'eances de l'Acad\'emie des sciences}, 157:708--710, 1913.

\bibitem{postSagnac1967}
E.~J. Post.
\newblock Sagnac {{Effect}}.
\newblock {\em Reviews of Modern Physics}, 39(2):475--493, April 1967.

\bibitem{anderson1994sagnac}
Ronald Anderson, HR~Bilger, and GE~Stedman.
\newblock {S}agnac effect: A century of {E}arth-rotated interferometers.
\newblock {\em American Journal of Physics}, 62(11):975--985, 1994.

\bibitem{malykin2000sagnac}
Grigorii~B Malykin.
\newblock The sagnac effect: correct and incorrect explanations.
\newblock {\em Physics-Uspekhi}, 43(12):1229, 2000.

\bibitem{barrett2014sagnac}
Brynle Barrett, R{\'e}my Geiger, Indranil Dutta, Matthieu Meunier, Benjamin Canuel, Alexandre Gauguet, Philippe Bouyer, and Arnaud Landragin.
\newblock The {S}agnac effect: 20 years of development in matter-wave interferometry.
\newblock {\em Comptes Rendus Physique}, 15(10):875--883, 2014.

\bibitem{Macek1963}
W.~M. Macek and D.~T.~M. Davis.
\newblock Rotation rate sensing with traveling-wave ring lasers.
\newblock {\em Applied Physics Letters}, 2(3):67, February 1963.

\bibitem{vali1976fiber}
Victor Vali and RW~Shorthill.
\newblock Fiber ring interferometer.
\newblock {\em Applied optics}, 15(5):1099--1100, 1976.

\bibitem{lefevre2014fiber}
Herv{\'e}~C Lef{\`e}vre.
\newblock The fiber-optic gyroscope, a century after {S}agnac's experiment: The ultimate rotation-sensing technology?
\newblock {\em Comptes Rendus Physique}, 15(10):851--858, 2014.

\bibitem{PhysRevLett.133.013601}
Angela D.~V. Di~Virgilio, Francesco Bajardi, Andrea Basti, Nicol\`o Beverini, Giorgio Carelli, Donatella Ciampini, Giuseppe Di~Somma, Francesco Fuso, Enrico Maccioni, Paolo Marsili, Antonello Ortolan, Alberto Porzio, and David Vitali.
\newblock Noise level of a ring laser gyroscope in the femto-rad/s range.
\newblock {\em Phys. Rev. Lett.}, 133:013601, Jul 2024.

\bibitem{bertocchi2006single}
Guillaume Bertocchi, Olivier Alibart, Daniel~Barry Ostrowsky, S{\'e}bastien Tanzilli, and Pascal Baldi.
\newblock Single-photon sagnac interferometer.
\newblock {\em Journal of Physics B: Atomic, Molecular and Optical Physics}, 39(5):1011, 2006.

\bibitem{finkExperimental2017}
Matthias Fink, Ana {Rodriguez-Aramendia}, Johannes Handsteiner, Abdul Ziarkash, Fabian Steinlechner, Thomas Scheidl, Ivette Fuentes, Jacques Pienaar, Timothy~C. Ralph, and Rupert Ursin.
\newblock Experimental test of photonic entanglement in accelerated reference frames.
\newblock {\em Nature Communications}, 8(1):15304, May 2017.

\bibitem{restuccia2019photon}
Sara Restuccia, Marko Toro{\v{s}}, Graham~M. Gibson, Hendrik Ulbricht, Daniele Faccio, and Miles~J. Padgett.
\newblock Photon bunching in a rotating reference frame.
\newblock {\em Phys. Rev. Lett.}, 123:110401, Sep 2019.

\bibitem{Fink_2019}
Matthias Fink, Fabian Steinlechner, Johannes Handsteiner, Jonathan~P Dowling, Thomas Scheidl, and Rupert Ursin.
\newblock Entanglement-enhanced optical gyroscope.
\newblock {\em New Journal of Physics}, 21(5):053010, may 2019.

\bibitem{silvestri2024experimental}
Raffaele Silvestri, Haocun Yu, Teodor Str\"{o}mberg, Christopher Hilweg, Robert~W. Peterson, and Philip Walther.
\newblock Experimental observation of {E}arth's rotation with quantum entanglement.
\newblock {\em Science Advances}, 10(24):eado0215, 2024.

\bibitem{torosRevealing2020}
Marko Toro{\v s}, Sara Restuccia, Graham~M. Gibson, Marion Cromb, Hendrik Ulbricht, Miles Padgett, and Daniele Faccio.
\newblock Revealing and concealing entanglement with noninertial motion.
\newblock {\em Physical Review A}, 101(4):043837, April 2020.

\bibitem{cromb2023mechanical}
Marion Cromb, Sara Restuccia, Graham~M. Gibson, Marko Toro{\v{s}}, Miles~J. Padgett, and Daniele Faccio.
\newblock Mechanical rotation modifies the manifestation of photon entanglement.
\newblock {\em Phys. Rev. Res.}, 5:L022005, Apr 2023.

\bibitem{bittermann2024non}
Julius~Arthur Bittermann, Matthias Fink, Marcus Huber, and Rupert Ursin.
\newblock Non-inertial motion dependent entangled {B}ell-state.
\newblock {\em arXiv preprint arXiv:2401.05186}, 2024.

\bibitem{torosGeneration2022}
Marko Toro\ifmmode~\check{s}\else \v{s}\fi{}, Marion Cromb, Mauro Paternostro, and Daniele Faccio.
\newblock Generation of entanglement from mechanical rotation.
\newblock {\em Phys. Rev. Lett.}, 129:260401, Dec 2022.

\bibitem{Kochen1967}
Simon Kochen and E.~Specker.
\newblock The problem of hidden variables in quantum mechanics.
\newblock {\em Indiana University Mathematics Journal}, 17(1):59, 1967.

\bibitem{van2005single}
Steven~J van Enk.
\newblock Single-particle entanglement.
\newblock {\em Physical Review A}, 72(6):064306, 2005.

\bibitem{adhikari2010swapping}
S~Adhikari, AS~Majumdar, Dipankar Home, and AK~Pan.
\newblock Swapping path-spin intraparticle entanglement onto spin-spin interparticle entanglement.
\newblock {\em Europhysics Letters}, 89(1):10005, 2010.

\bibitem{kumari2019swapping}
Asmita Kumari, Abhishek Ghosh, Mohit~Lal Bera, and AK~Pan.
\newblock Swapping intraphoton entanglement to interphoton entanglement using linear optical devices.
\newblock {\em Physical Review A}, 99(3):032118, 2019.

\bibitem{bose2017spin}
Sougato Bose, Anupam Mazumdar, Gavin~W Morley, Hendrik Ulbricht, Marko Toro{\v{s}}, Mauro Paternostro, Andrew~A Geraci, Peter~F Barker, MS~Kim, and Gerard Milburn.
\newblock Spin entanglement witness for quantum gravity.
\newblock {\em Physical review letters}, 119(24):240401, 2017.

\bibitem{marletto2017gravitationally}
Chiara Marletto and Vlatko Vedral.
\newblock Gravitationally induced entanglement between two massive particles is sufficient evidence of quantum effects in gravity.
\newblock {\em Physical review letters}, 119(24):240402, 2017.

\bibitem{polino2024photonic}
Emanuele Polino, Beatrice Polacchi, Davide Poderini, Iris Agresti, Gonzalo Carvacho, Fabio Sciarrino, Andrea Di~Biagio, Carlo Rovelli, and Marios Christodoulou.
\newblock Photonic implementation of quantum gravity simulator.
\newblock {\em Advanced Photonics Nexus}, 3(3):036011--036011, 2024.

\bibitem{krisnandaRevealing2017}
Tanjung Krisnanda, Margherita Zuppardo, Mauro Paternostro, and Tomasz Paterek.
\newblock Revealing {{Nonclassicality}} of {{Inaccessible Objects}}.
\newblock {\em Physical Review Letters}, 119(12):120402, September 2017.

\bibitem{krisnandaPhotosyntesis2018}
Tanjung Krisnanda, Chiara Marletto, Vlatko Vedral, Mauro Paternostro, and Tomasz Paterek.
\newblock Probing quantum features of photosynthetic organisms.
\newblock {\em npj Quantum Inf}, 4(60), November 2018.

\bibitem{clauser1969proposed}
John~F Clauser, Michael~A Horne, Abner Shimony, and Richard~A Holt.
\newblock Proposed experiment to test local hidden-variable theories.
\newblock {\em Physical review letters}, 23(15):880, 1969.

\bibitem{cirel1980quantum}
Boris~S Cirel'son.
\newblock Quantum generalizations of {B}ell's inequality.
\newblock {\em Letters in Mathematical Physics}, 4:93--100, 1980.

\bibitem{Popescu1997}
Sandu Popescu, Lucien Hardy, and Marek \ifmmode~\dot{Z}\else \.{Z}\fi{}ukowski.
\newblock Revisiting bell's theorem for a class of down-conversion experiments.
\newblock {\em Phys. Rev. A}, 56:R4353--R4356, Dec 1997.

\bibitem{post1967sagnac}
Evert~Jan Post.
\newblock Sagnac effect.
\newblock {\em Reviews of Modern Physics}, 39(2):475, 1967.

\bibitem{korsbakken2004fulling}
Jan~Ivar Korsbakken and Jon~Magne Leinaas.
\newblock Fulling-{U}nruh effect in general stationary accelerated frames.
\newblock {\em Physical Review D}, 70(8):084016, 2004.

\bibitem{zych2012general}
Magdalena Zych, Fabio Costa, Igor Pikovski, Timothy~C Ralph, and {\v{C}}aslav Brukner.
\newblock General relativistic effects in quantum interference of photons.
\newblock {\em Classical and Quantum Gravity}, 29(22):224010, 2012.

\bibitem{brady2021frame}
Anthony~J Brady and Stav Haldar.
\newblock Frame dragging and the hong-ou-mandel dip: Gravitational effects in multiphoton interference.
\newblock {\em Physical Review Research}, 3(2):023024, 2021.

\bibitem{HOM_Kerr_2022}
S.~P. Kish and T.~C. Ralph.
\newblock Quantum effects in rotating reference frames.
\newblock {\em AVS Quantum Science}, 4(1):011401, 2022.

\bibitem{barzelObserver2022}
Roy Barzel, David~Edward Bruschi, Andreas~W. Schell, and Claus L{\"a}mmerzahl.
\newblock Observer dependence of photon bunching: {{The}} influence of the relativistic redshift on {{Hong-Ou-Mandel}} interference.
\newblock {\em Physical Review D}, 105(10):105016, May 2022.

\bibitem{aharonov1984quantum}
Yakir Aharonov and Tirzah Kaufherr.
\newblock Quantum frames of reference.
\newblock {\em Physical Review D}, 30(2):368, 1984.

\bibitem{giacomini2019relativistic}
Flaminia Giacomini, Esteban Castro-Ruiz, and {\v{C}}aslav Brukner.
\newblock Relativistic quantum reference frames: the operational meaning of spin.
\newblock {\em Physical review letters}, 123(9):090404, 2019.

\bibitem{Brunner2007}
Nicolas Brunner, Nicolas Gisin, Valerio Scarani, and Christoph Simon.
\newblock Detection loophole in asymmetric bell experiments.
\newblock {\em Phys. Rev. Lett.}, 98:220403, May 2007.

\end{thebibliography}

\end{document}